# Evolution of Cooperation in an Incentive Based Business Game Environment

Sanat Kumar Bista[1], Keshav P Dahal[1], Peter I Cowling[1]

[1]MOSAIC Research Centre, University of Bradford, Bradford, West Yorkshire, BD7 1DP, UK

**Abstract**

This paper discusses our investigation into the evolution of cooperative players in an online business environment. We explain our design of an incentive based system with its foundation over binary reputation system whose proportion of reward or punishment is a function of transaction value and the player's past history of cooperation. We compare the evolution of cooperation in our setting with non-incentive based environment and our findings show that the incentive based method is more suitable for the evolution of trustworthy players.

## 1  Introduction

Trust is a crucial component of society. As a foundation of human civilization, trust continues to be important in all aspects of life. Whether we rely or not on some object is guided by how trust worthy we believe it to be. In a social context, trust worthiness is assessed in several ways, for example by referring to the past history of interaction, word-of-mouth, reliable third party certification, social reputation etc [1-3].

Computational model of trust in online societies is not straightforward. One of the common ways of assessing trustworthiness in online societies is a reputation mechanism and it has emerged as an important component of electronic markets in eliciting cooperation within loosely coupled and geographically dispersed economic agents [4]. Online auction and business sites like eBay, Yahoo Auction, Amazon.com etc use simple yet effective reputation management frameworks to provide their users with reputation information. The success of these trading environments demonstrates that reputation mechanisms are an effective way of inferring trust worthiness in the transacting parties. However, with strategic players in application fronts it has become increasingly difficult to identify a trustworthy partner for interaction. These types of systems largely preserve user anonymity and this brings additional challenges. Thus it is necessary to identify the parameters that would contribute positively in making possible the evolution of a cooperative society. The system we discuss and propose is a possible step towards this.

## 2   Background

The trust and reputation management framework we are considering builds upon the requirements of online market places. Online reputation systems like that of eBay represents a simple and successful binary reputation based *quality of service* monitoring utility the in existing online business environments. The problem for the eBay reputation system [5] is to compute a trustworthiness value given the total number of positive feedbacks ($V_p$) and the total number of negative feedbacks ($V_n$) received by the player (the term *player* would be used to describe the buyer as well as seller). If any player **ABC** has a total of 4746 unique positive feed backs ($V_p$) and a total of 9 unique negative feedbacks ($V_n$). The positive feedback value expressed in percentage is thus the ratio of positive votes to the sum of positive and negative votes, which in this case becomes 99.8%.

The rating process however does not consider the value of good that is being sold or purchased and the reputation of the player who is a seller or buyer in the process. This information is quite significant while assessing the quality of feedback and reputation consequently. It is always possible that a player might build a good reputation score by transacting small valued goods at first and later on might 'cheat' in a high value transaction [6]. Similarly, the feedback provided by a player with high reputation should be more meaningful than that provided by players with lower reputation; these points seem to be neglected by the current online recommendation systems. In our investigation, we include these parameters and present a comparative analysis of the results of evolution that was obtained by considering and not considering these parameters. We show that the inclusion of these parameters in the business process contributes in the evolution of a cooperative society with a larger number of cooperative players in it.

The experiments for this investigation were carried out in an Iterated Prisoner's Dilemma [7] like setting over a spatial distribution of players. IPD environment represents social dilemma situation [8] [7] [9]. A typical online business setting has a dilemma situation, as a buyer doesn't know the seller and either of the parties are not sure of cooperation. Defection in a one shot business interaction seems to be attractive but in a repeated interaction, cooperation might still be attractive as it gains an increased reputation for the player and this could be helpful in future business, hence the dilemma.  The pay offs for Temptation(T), Reward(R), Punishment(P) and Sucker(S) within an Iterated Prisoner's dilemma game would strictly follow the following two inequalities, (i). T>R>P>S, and (ii). 2R>T+S [7]. The pay offs in a typical business game would have a different relationship (as expressed in Table 2 Section 4).  The real difference that this has brought about is in making defection even more attractive over cooperation as the reward value now becomes equal to punishment. This endangers the dilemma situation and a right incentive for cooperation needs to be added for the dilemma to continue.

In the simulation, number of players would play the cooperation-defection over generations in a genetic algorithm based environment. The payoff values as obtained would work as a fitness function for the GA based simulation, where the player strategies are represented by the chromosomes and each chromosome is a fixed length representation of player strategies in terms of cooperation (C) and defection (D). Essentially, the system searches for an optimal strategy.

A memory-3 game with four specific possible moves (CC, CD, DC, DD) is played between the players thus making the chromosome size 64 ($4^3$). The original IPD tournaments were described by Axelrod in [7] and it was first programmed by Forrest [10]. Axelrod used additional 6 bits to determine the first three moves. A variation of this was used by Errity in [11] and we are following the same scheme of additional bit encoding, in which 7 extra bits are used for encoding actions for the first three relative moves (relative to opponent moves). In this approach it is not required to encode an assumption of the pre-game history [11]. For reproduction, the system is capable of performing a crossover as well as mutation. During crossover, both the parent chromosomes are broken in at the same random point. Values of 0.001 for the mutation probability and 0.5 for crossover probability have been used throughout the game. The players have been categorized into six different types as defined by the percentage of cooperative actions exhibited in their strategy as represented in their chromosomes. Table 1 below presents the classification:

**Table 1** Player Classification

| Player Type | Cooperation (%) |
|---|---|
| Very Cooperative | > 65 |
| Cooperative | 55 to 65 |
| Good | 50 to 54 |
| Okay | 45 to 49 |
| Dishonest | 35 to 44 |
| Very Dishonest | < 35 |

## 3   Related Works

Trust and Reputation in the context of e-commerce environments and Peer to Peer systems have been an area of significant interest in recent. Aberer in [8] outlines the complexity of Trust and Reputation and discusses different approaches to computing trust and reputation. The authors have considered evolutionary approach as one of the many popular approaches that game theorists have been using. In [12] the authors have presented a social mechanism of reputation management in electronic communities. In their discussion around electronic communities, the authors have described the Prisoner's dilemma situation in it. Our choice of a Prisoner's Dilemma like environment to represent online business interaction has been justified by their discussion in these papers.

In [6] the authors describe their design of a reputation management system for peer-to-peer systems in electronic communities. While listing the problems of electronic communities, the authors have marked the lack of incentives in rating to be a major one. In addition the paper also highlights the existing systems' lack of ability in handling strategic players. Our research is fueled by these two listings.

In a related work Janssen in [9] has studied the role of reputation scores in the evolution of cooperation in online e-commerce sites. The author discusses whether or not reputation alone can be meaningful in evolving a cooperative society. The paper concludes that high level cooperation is not only possible with reputation scores. The author investigates the work in a one-shot prisoner's dilemma like environment.

Based on these findings, our work in this paper concentrates on investigating the possible role of incentive in the evolution of cooperation.

## 4  Problem Definition and Incentive Based Model

The problem we are considering is a typical business game between a buyer and a seller. The corresponding actions and the pay offs are explained by the matrix representation in table 2 below:

**Table 2** Pay-off Matrix for business game

$P_{buyer}$

| | | Cooperate (C) | Defect (D) |
|---|---|---|---|
| $P_{seller}$ | C | $R_{seller}$ = Money <br> $R_{buyer}$ = Good(s) | $S_{seller}$ = - (Money) <br> $T_{Buyer}$ = Money + Good(s) |
| | D | $T_{seller}$ = Good(s) + Money <br> $S_{Buyer}$ = - (Good(s)) | $P_{seller}$ = Money <br> $P_{buyer}$ = Good (s) |

Here, R represents the reward pay off, S the suckers pay off, T for Temptation payoff and P for punishment payoff. It is clear from the table that in any case the player is in safe side to play a defection. If the other player cooperates it is going to receive a Temptation pay off which is twice the amount of Reward (as the defector would have goods as well as money in his hand), and even if the other player plays a defection it is going to receive a Punishment payoff which is equal in value to the reward pay off (as the player would still have in his hand either the goods or money). A player who cooperated while his opponent defected will loose money as well as goods. If we think of any online business environment *preserving total anonymity* of players, then the situation is closely resembled by the one described above. As an example, if we keep the 'physical' means of user identification and loss compensation schemes as a constant, this situation reflects the eBay like business scenario. This situation should not exist as it might result in a high number of selfish players in the society, a fact that is demonstrated in our experimental results.

To avoid this situation we focus our investigation on what impact the inclusion of player reputation and price related data can be in the evolution of cooperation. In our model we use '*bonus reward*' as an incentive to cooperative behavior in the game. Mutual cooperation in a game representing a single transaction would result in a payoff equivalent to the reward as in Table 2. Plus, we would assign a bonus reward, computed as a function of the player reputation and value of the goods. In the other hand, in a case where both parties defect each other, their corresponding pay off values are subjected to a decrement in pay off indicating a more severe penalty for punishment. In the later case bonus reward would be subtracted from the reward value. In a simple approach towards this we base the reward and punishment pay off to be dependent on the following two parameters:

    a. The price value of the transaction(equal to the Reward for cooperation )
    b. The existing cooperation probability (reputation) of the player as given by its history of cooperation and defection.

The corresponding actions and associated pay offs for an incentive based setting is represented by Table 3 below:

**Table 3** Incentive compatible Pay-off Matrix for business game

|  |  | $P_{buyer}$ | |
|---|---|---|---|
|  |  | **Cooperate (C)** | **Defect (D)** |
| $P_{seller}$ | **C** | $R_{seller} = \text{Val}_G + \theta_{Pseller} \times \text{Val}_G$ <br> $R_{buyer} = \text{Val}_G + \theta_{PBuyer} \times \text{Val}_G$ | $s_{seller} = -Val_G$ <br> $T_{Buyer} = 2 \times Val_G$ |
|  | **D** | $T_{seller} = 2 \times Val_G$ <br> $s_{Buyer} = -Val_G$ | $P_{seller} = \text{Val}_G - (\theta_{Pseller} \times \text{Val}_G)$ <br> $P_{buyer} = \text{Val}_G - (\theta_{PBuyer} \times \text{Val}_G)$ |

Here, **Val$_G$** represents the value of goods being transacted and $\theta$ represents the reputation of the player. The reputation information is maintained by the system as a vector with a total number of Cooperation (C) and Defection (D):

$$T_H = \begin{bmatrix} C \\ D \end{bmatrix} \quad (1)$$

The expected probability of cooperation is given by the following expression:

$$E(P_n) = \frac{C}{C+D} \quad (2)$$

Where, the values for C and D are derived from the transaction history in (1).

## 5  Experimental Setup and Results

The experiments were carried out in two phases. In the first phase a total of 2500 players were selected to play a non-incentive business game for 5000 generations with 100 iterations in each generation. The pay-off values for this game were based on the explanation provided in Table 2 above. In the second phase by keeping the other parameters same, a pro-incentive business model with pay-off values as listed in Table 3 above was simulated. The system recorded the readings of the player evolution and the cooperative and defective moves in each interaction. The results obtained are an average of 10 rounds of simulation. We assume that each player initially has a truth telling probability of 1. Further, we assume that the players are transacting goods of same value through out.

The stacked bar diagrams in figure 1 below shows the percentage growth share in the evolution of the six different types of players in each of the settings. In the diagram, it is clearly shown that the population of cooperative players (Very cooperative, cooperative, good and okay players) rise to higher values as the evolution continues in a pro-incentive setting. Classification of these players in terms of probability of cooperation was presented in Table 1 before. The population of non cooperative players (very dishonest and dishonest players) is high and continues to grow in a non-incentive setting.

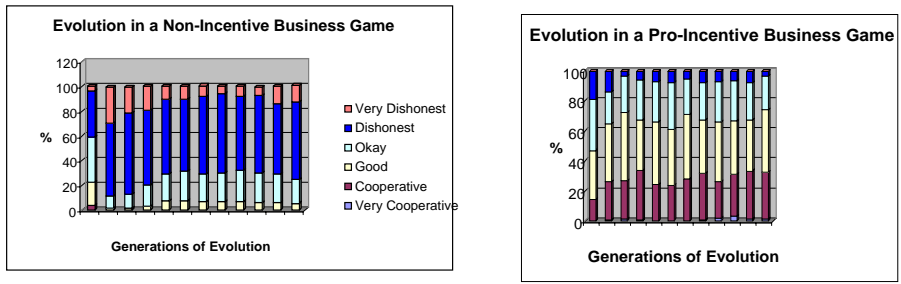

**Fig. 1.** Evolution of different player types over 5000 generations (1,100,500,1000,1500,2000,2500,3000,3500,4000,4500 and 5000 respectively) in a non-incentive and pro-incentive business game environment

The evolution trends of each type of player were observed in comparison with respect to each game setting. The graphs in Fig.2 below represent the comparative trend of evolution of the six different types of players. The evolution trend in general shows that the pro-incentive setting is favorable for the evolution of cooperative players while the non-incentive setting is favorable to the non-cooperative player evolution.

Evolution of Cooperation in an Incentive Based Business Game Environment

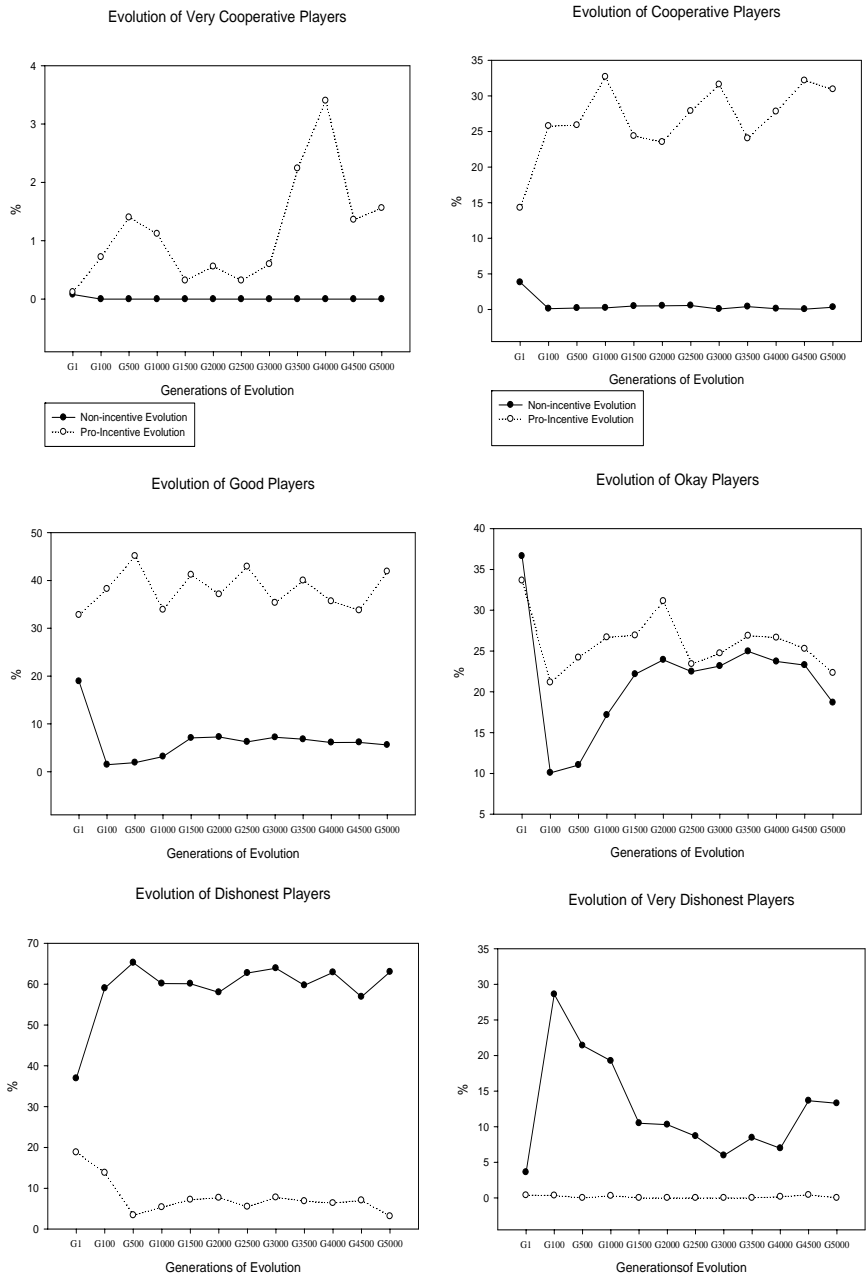

Fig.2. Comparative evolution of six different player types in a non-incentive and pro-incentive business game setting

Another interesting aspect is the reputation of the players. In each play in the game, the cooperative and defective moves of the players were recorded in order to calculate the total reputation score. Reputation would be calculated as in expression (2) above. An average reputation score of 0.98 was recorded for pro-incentive business game whereas a very low average score of 0.003 was recorded for non-incentive setting. This result is in correlation to the population of non-cooperative players who would defect in maximum as opposed to the cooperative instances of good players in the pro-incentive environment.

## 6   Discussion and Conclusion

"Hard security mechanisms" like authentication, access control, encryption etc have been in use in different online business environments to reduce the chances of fraudulent acts [13, 14]. Such mechanisms might also include the registration requirements, requirements of personal details including bank and physical address, telephone numbers etc. While these mechanisms do certainly contribute in reducing possibly fraudulent players from appearing in the market, they also reduce the level of participation in terms of numbers. Further, dishonest behaviors can also be demonstrated by players who pass it. The notion of trustworthiness in online societies is really complex [8], and such hard security mechanisms might not be enough to curb on the temptation of defectors.

In an eBay like online business environment, if this was completely open, meaning that there would be completely no registration requirements for players and that the system would preserve total anonymity, the situation is in the worst case similar to the one depicted in our non-incentive business game. We suggested a pro-incentive model which demonstrated to be favorable for the evolution of cooperative players in society thus, leading to cooperation and trustworthiness with highly reputed players in it. Our investigation shows that the pro-incentive model which is an interrelated representation of cooperative behavior and reputation would be even more suitable for an open business environment. Our future investigation in this line could involve the formalization of incentive model, specifying the reputation to reflect the incentive for cooperation.